\documentclass[physics,article,accept,pdftex,moreauthors]{Definitions/mdpi}
\usepackage{amsmath}
\usepackage{amsfonts}
\usepackage{amssymb,bm}
\usepackage{siunitx}
\usepackage{color}
\usepackage{braket}
\usepackage{graphics}

\firstpage{952}
\pubvolume{5}
\issuenum{4}
\articlenumber{062}
\pubyear{2023}
\copyrightyear{2023}
\datereceived{24 July 2023 }
\daterevised{24 August 2023 } 
\dateaccepted{31 August 2023 }
\datepublished{27 September 2023 }
\hreflink{https://doi.org/10.3390/physics5040062} 

\Title{Casimir Effect Invalidates the Drude Model for Transverse Electric
Evanescent Waves}

\TitleCitation{Casimir Effect Invalidates the Drude Model for Transverse Electric
Evanescent Waves}

\Author{{Galina L. Klimchitskaya} ${}^{1,2}$*\orcidA{}
and
Vladimir M. Mostepanenko ${}^{1,2,3}$\orcidB{}}

\AuthorNames{Galina L. Klimchitskaya,
Vladimir M. Mostepanenko}

\AuthorCitation{Klimchitskaya, G.L.;  Mostepanenko, V.M.}

\address{%
$^{1}$ \quad Central Astronomical Observatory at Pulkovo of the Russian Academy of Sciences, 196140 Saint Petersburg, Russia\\
$^{2}$ \quad Peter the Great Saint Petersburg
Polytechnic University, 195251 Saint Petersburg, Russia\\
$^{3}$ \quad  {Kazan Federal University}, 420008 Kazan, Russia}

\corres{Correspondence: g.klimchitskaya@gmail.com}

\abstract{We consider the Casimir pressure between two metallic plates and
calculate the four contributions to it determined by the propagating
and evanescent waves and by the transverse magnetic and transverse
electric polarizations of the electromagnetic field. The range of
interplate separations is considered where nearly the whole pressure
has its origin in the electromagnetic response of conduction electrons.
In the Casimir physics, this response is described either by the
dissipative Drude model resulting in contradictions with the measurement
data or by the experimentally consistent but dissipationless plasma
model. It is shown that the total transverse magnetic contribution to
the Casimir pressure due to both the propagating and evanescent waves
and the transverse electric contribution due to only the propagating
waves, computed by means of the Drude model, correlate well with the
corresponding results obtained using the plasma model. The conclusion
is made that a disagreement between the theoretical predictions obtained
using the Drude model and precision measurements of the Casimir force
is not caused by the account of dissipation in itself, but arises from
an incorrect description of the response of metals to the low-frequency
transverse electric evanescent waves by this model. It is demonstrated
that the Drude model has no supporting experimental evidence in the range
of transverse electric evanescent waves, so that the above conclusion is
consistent with all available information. The alternative test of the
Drude model for the transverse electric evanescent waves suggested in
the framework of classical electrodynamics is discussed.}


\begin{document}

\section{Introduction}

The Casimir effect is the relativistic and quantum phenomenon which has
attracted widespread attention in the 75 years since its prediction in
1948 \cite{1}. This effect is very popular owing to its unusual character.
Casimir predicted that two parallel uncharged ideal metal planes at zero
temperature attract each other with the force which depends only on the
interplate separation and the fundamental constants $\hbar$ and $c$.
In 1955, Lifshitz demonstrated \cite{2} that the Casimir force falls into
the general theory of dispersion forces, which act between any material
bodies. From the point of view of the Lifshitz theory, both the van der
Waals and and Casimir forces are the manifestations of a single dispersion
force, but in different regions of separations and temperatures. The
Lifshitz theory makes it possible to calculate the Casimir force between
two thick material plates by using the response functions of plate
materials to the electromagnetic field in the form of frequency-dependent
dielectric permittivities.

The Casimir force is unique in being important for such diverse fields of
physics as theory of elementary particles, gravitation and cosmology,
quantum electrodynamics, condensed matter physics, atomic physics, and also
for nanotechnology. As a result, a great number of papers was devoted to
this subject during the last decades (see the lists of references in the
monographs \cite{3,4,5,6,7,8,9,10,11,12}). In doing so, much attention has
been paid to precision measurements of the Casimir force.

The present stage in measuring the Casimir force started with an experiment
\cite{13}, which used the configuration of an Au-coated spherical lense of
centimeter-size radius above an Au-coated plate. As was understood later,
the presence of the so-called patch potentials \cite{14} and surface
imperfections \cite{15} on the centimeter-size surfaces prevents from
reaching the highly precise results in measuring the Casimir force. The
highly accurate measurements were performed between a microscopic sphere
and a plate by means of an atomic force microscope and a micromechanical
torsional oscillator pioneered in \cite{16} and \cite{17}, respectively.

The many times repeated measurements of the Casimir force performed by
means of a micromechanical torsional oscillator \cite{18,19,20,21,22,23}
and an atomic force microscope \cite{24,25,26,27,28,29,30} led to the
unexpected result. It was found that the measurement data are in a
very good agreement with theoretical predictions of the Lifshitz theory
if the low-frequency response of metals to the electromagnetic field is
described by the dissipationless plasma model. If the dissipative Drude
model is used, which should describe the conduction electrons correctly,
the theoretical predictions are excluded by the data with certainty
\cite{18,19,20,21,22,23,24,25,26,27,28,29,30}. Only in a single experiment
\cite{31}, the force values computed by means of the Drude model were
confirmed, but the measurements were performed by means of a
centimeter-size spherical lens. As a result, the theoretical uncertainty
{ {due to patch potentials} }
removed by means of the fitting procedure exceeded the Casimir force
value by an order of magnitude.
{{ Moreover, the surface imperfections, which are always present on
lens surfaces, were not taken into account in this experiment}} \cite{15,31a}.

The contradiction between theoretical predictions of the Lifshitz theory
obtained using the apparently well-tested Drude model and measurements of
the Casimir force is often named the Casimir puzzle \cite{32,33,34}. A
rich variety of approaches has been suggested in the literature in an
effort to resolve it. One could mention an employment of the alternative
sets of the optical data \cite{35,36}, modeling the patch effect
\cite{14,31,37}, a more accurate account of the surface roughness
\cite{38,39,40}, refined theory for the sphere-plate geometry
\cite{41,42,43,44,45,46} etc. (see \cite{11,47,48,49} for a review).

Particular emphasis has been placed on the frequency region of the
anomalous skin effect where the Drude dielectric function becomes
inapplicable due to the spatial nonlocality \cite{50,51,52}. It was
found, however, that the corresponding correction to the Casimir
force is too small and cannot explain the discrepancy between the
measurement data and theory which uses the Drude model \cite{50}.

The important step was made in \cite{53,54} where it was shown that
large thermal correction to the Casimir force predicted by the Drude
model arises from the transverse electric \\ (s-polarized) evanescent waves
with low frequencies. This result was obtained by analyzing the
frequency spectrum of the thermal correction along the real frequency
axis. The predicted large thermal correction to the Casimir force, which
distinguishes the Drude model from the plasma model and the model of an
ideal metal, was also interpreted as arising from the contribution of
eddy (Foucault) current modes \cite{55,56}.

Furthermore, it was shown that at separations exceeding the thermal length
(i.e., above approximately 6 $\upmu$m at $T=300$K) the contributions of the
transverse electric propagating and evanescent waves to the total Casimir
force calculated using the Drude model are equal in magnitude and cancel
each other \cite{57}. According to \cite{58}, at large separations the
contributions of the transverse magnetic (p-polarized) and transverse
electric propagating waves are equal regardless of what dielectric model
(Drude or plasma) is used in computations. As to the contribution of
transverse magnetic evanescent waves, it is equal to zero for both the
Drude and plasma models. Thus, at large separations, the difference in
Casimir forces computed using the Drude and plasma models originates
solely from the contribution of transerse electric evanescent waves.

In this paper, we investigate the contributions of both the transverse
magnetic and transverse electric propagating and evanescent waves into
the Casimir force per unit area (i.e., the Casimir pressure) for two
parallel Au-coated plates in the experimentally relevant separation region
from 0.5 to 4 $\upmu$m where the total
force value, in the limits of measurement errors, is determined by the
dielectric response of conduction electrons. The contributions of the
transverse magnetic and transverse electric propagating and evanescent
waves are calculated in the framework of the Lifshitz theory employing
either the Drude or the plasma model. For this purpose, we combine the
computational results obtained using the formalisms represented in terms
of the pure imaginary and real frequencies.

It is shown that the contributions of the transverse magnetic waves to
the total Casimir force computed using the Drude and plasma models
nearly coincide. The contributions of the transverse electric propagating
waves to the Casimir force computed using the Drude and plasma models
also turned out to be rather close. As a result, the relatively big
difference between the theoretical predictions for the total Casimir force
made by means of the Drude and plasma models over the experimentally
relevant range of separations comes from different contributions of the
transverse electric evanescent waves. Taking into account that this big
difference is experimentally excluded by the measurement data of numerous
experiments mentioned above, the conclusion is made that the Drude model
breaks down in the region of transverse electric evanescent waves.
We demonstrate that this conclusion is not in conflict with numerous
experimental tests of the Drude model. The obtained results are discussed
in connection with the role of dissipation of conduction electrons in the
Lifshitz theory.

The paper is organized as follows. In Section 2, we briefly present the
formalisms of the Lifshitz theory in terms of either pure imaginary or
real frequencies separating the contributions of the transverse magnetic
and transverse electric polarizations and the propagating and evanescent
waves.  Section 3 is devoted to computations of the Casimir pressure between
metallic plates using the Drude and the plasma models and the optical data
for the complex index of refraction. In Section 4, the contributions of
the propagating and evanescent waves are studied for the transverse magnetic
and transverse electric polarizations using the Drude and plasma models.
Section 5 discusses the failure of the Drude model for the transverse
electric evanescent waves, the role of dissipation of conduction electrons,
and the possibilities of alternative tests disconnected with the Casimir
effect. Section 6 contains our conclusions.

\section{Formalisms of the Lifshitz Theory in Terms of Real or Pure
Imaginary Frequencies}
\newcommand{\nrmP}{P_{\rm D}^{\,{0}}}
\newcommand{\ve}{{\varepsilon}}
\newcommand{\fve}{{\varepsilon(\omega)}}
\newcommand{\kb}{{k_{\bot}}}
\newcommand{\skb}{{k_{\bot}^2}}
\newcommand{\okb}{{(\omega,k_{\bot})}}

We consider the Casimir force per unit area of two similar metallic plates
described by the dielectric permittivity $\fve$, i.e., the Casimir pressure.
The plates are at temperature $T$ in thermal equilibrium with the environment
and are separated by a distance $a$. Then, the Casimir pressure can be
expressed by the Lifshitz formula \cite{2}. This formula can be presented
in terms of real frequencies or pure imaginary (Matsubara) frequencies.

In terms of real frequencies, the Casimir pressure is given by the sum of
contributions from the propagating and evanescent waves each of which, in its
turn, consists of two components determined by the transverse magnetic
(TM) and transverse electric (TE) polarizations

\begin{equation}
P(a,T)=P_{\rm TM}^{\,\rm prop}(a,T)+P_{\rm TE}^{\,\rm prop}(a,T)+
P_{\rm TM}^{\,\rm evan}(a,T)+P_{\rm TE}^{\,\rm evan}(a,T).
\label{eq1}
\end{equation}

\noindent
Here \cite{11}

\begin{equation}
P_{\rm TM,TE}^{\,\rm prop}(a,T)=-\frac{\hbar}{2\pi^2}\int\limits_{0}^{\infty}\!\!d\omega
\coth\frac{\hbar\omega}{2k_BT}
\int\limits_0^{\omega/c}\!\!d\kb\kb {\rm Im}\left[q
\frac{r_{\rm TM,TE}^2\okb e^{-2aq}}{1-r_{\rm TM,TE}^2\okb e^{-2aq}}\right]
\label{eq2}
\end{equation}

\noindent
and

\begin{equation}
P_{\rm TM,TE}^{\,\rm evan}(a,T)=-\frac{\hbar}{2\pi^2}\int\limits_{0}^{\infty}\!\!d\omega
\coth\frac{\hbar\omega}{2k_BT}
\int\limits_{\omega/c}^{\infty}\!\!d\kb\kb q\,{\rm Im}
\frac{r_{\rm TM,TE}^2\okb e^{-2aq}}{1-r_{\rm TM,TE}^2\okb e^{-2aq}}.
\label{eq3}
\end{equation}

In these equations, the following notations are introduced. The Boltzmann constant
is $k_B$, the magnitude of the wave vector projection on the plane of plates is
$\kb$, the reflection coefficients for the TM and TE polarizations are

\begin{equation}
r_{\rm TM}\okb=\frac{\fve q-p}{\fve q+p},
\qquad
r_{\rm TE}\okb=\frac{q-p}{q+p},
\label{eq4}
\end{equation}

\noindent
and

\begin{equation}
q\equiv q\okb=\left(\skb-\frac{\omega^2}{c^2}\right)^{1/2},
\qquad
p\equiv p\okb=\left[\skb-\fve\frac{\omega^2}{c^2}\right]^{1/2}.
\label{eq5}
\end{equation}

{ {Note that by solving the Maxwell equations with the continuity boundary
conditions on the surfaces of metallic plates, one determines the Casimir
energy via the sum  of discrete photon eigenfrequencies (or the cavity modes
or the wave guide modes, as they are often referred to}} \cite{58a}).
The continuous frequencies in Equation (3) appear after performing
a summation over the discrete frequencies by means of the argument principle.

As is seen from (\ref{eq2}), for the propagating waves $\kb\leqslant\omega/c$
in accordance to the mass-shell equation in free space. The quantity $q$ in
this case is pure imaginary and the integrand in (\ref{eq2}) contains the
rapidly oscillating function $\exp(-2aq)$ that plagues numerical computations.
For the evanescent waves in (\ref{eq3}), the mass-shell equation is violated
$\kb>\omega/c$, but the quantity $q$ takes real values making accessible
computations of $P_{\rm TM,TE}^{\rm evan}$ by means of (\ref{eq3}).

One can conclude that equations (\ref{eq1})--(\ref{eq3}) are not convenient
for computations of the total Casimir pressure (\ref{eq1}), but the
contributions $P_{\rm TM,TE}^{\rm evan}$ from the evanescent waves can be
 computed by (\ref{eq3}).

In terms of the pure imaginary Matsubara frequencies,
$\omega=i\xi_l=2\pi i k_BTl/\hbar$ with $l=0,\,1,\,2,\,\ldots$, the Casimir
pressure is expressed by the most commonly used Lifshitz formula

\begin{equation}
P(a,T)=P_{\rm TM}(a,T)+P_{\rm TE}(a,T),
\label{eq6}
\end{equation}

\noindent
where \cite{11}

\begin{equation}
P_{\rm TM,TE}(a,T)=-\frac{k_BT}{\pi}\sum_{l=0}^{\infty}{\vphantom{\sum}}^{\!\prime}
\!\int\limits_0^{\infty}\!\kb d\kb q_l
\frac{r_{\rm TM,TE}^2(i\xi_l,\kb)e^{-2aq_l}}{1-r_{\rm TM,TE}^2(i\xi_l,\kb)e^{-2aq_l}}.
\label{eq7}
\end{equation}

The prime on the summation sign in (\ref{eq7}) divides the terms with $l=0$ by 2,
and the reflection coefficients are again defined by (\ref{eq4}) with $\omega=i\xi_l$,
so that in line with (\ref{eq5})

\begin{equation}
q=q_l\equiv q(i\xi_l,\kb)=\left(\skb+\frac{\xi_l^2}{c^2}\right)^{1/2},
\qquad
p=p_l\equiv p(i\xi_l,\kb)=\left(\skb+\ve_l\frac{\xi_l^2}{c^2}\right)^{1/2},
\label{eq8}
\end{equation}

\noindent
where $\ve=\ve_l\equiv\ve(i\xi_l)$

Equation (\ref{eq7}) is convenient for numerical computations of $P_{\rm TM,TE}$,
but it alone does not allow computation of the contributions from the propagating
and evanescent waves. In fact all the four components of the Casimir pressure on
the right-hand side of (\ref{eq1}) can be found by the combined application
of the Lifshitz formulas (\ref{eq3}) in terms of real frequencies  and (\ref{eq7})
in terms of the Matsubara frequencies. For this purpose, it is necessary to
compute the contributions  $P_{\rm TM,TE}^{\rm evan}$ by (\ref{eq3}) and the total
Casimir pressures $P_{\rm TM,TE}$ by (\ref{eq7}). Then, the remaining contributions
$P_{\rm TM,TE}^{\rm prop}$ are found from

\begin{equation}
P_{\rm TM,TE}^{\,\rm prop}(a,T)=P_{\rm TM,TE}(a,T)-
P_{\rm TM,TE}^{\,\rm evan}(a,T).
\label{eq9}
\end{equation}

The numerical computations of all four components of the total Casimir pressure
between metallic plates using different dielectric functions of a metal are
presented in the next sections.

\section{Calculation of the Casimir Pressure Between Metalic Plates Using the
Drude and Plasma Models}

It has been known that the dielectric response of metals to the electromagnetic
field is determined by the combined action of conduction and bound (core)
electrons. In doing so, the corresponding contributions to the dielectric
permittivity make a substantially different impact on the Casimir pressure \cite{11}.
At short separations between the plates (up to tens of nanometers), the major
contribution to the Casimir pressure is given by the region of very high
frequencies, where $\ve$ is fully determined by the core electrons. In the
transition region (from tens to hundreds of nanometers), both the conduction and
core electrons determine the value of $\ve$ at the frequencies contributing to the
Casimir pressure. Finally, at separations exceeding several hundreds of
nanometers, only the conduction electrons determine the dielectric response of
metals at the characteristic (low) frequencies.

Taking into account that the problem of disagreement between experiment and
theory discussed in Section~1 arises exclusively due to the role of conduction
electrons, it is appropriate to consider the separation region where the role
of core electrons in computations of the Casimir pressure is negligibly small.
In this section, the sought for region is found for two Au plates at room
temperature $T=300\,$K (the same results are valid for the plates made of any
material coated with an Au layer of thickness exceeding several tens of
nanometers \cite{11}).

As discussed in Section~1, the conduction electrons are commonly described
by the dielectric permittivity of the dissipative Drude model

\begin{equation}
\ve_{\rm D}(\omega)=1-\frac{\omega_p^2}{\omega(\omega+i\gamma)}, \qquad
\ve_{{\rm D},l}=1+\frac{\omega_p^2}{\xi_l(\xi_l+\gamma)},
\label{eq10}
\end{equation}

\noindent
where, for Au, the plasma frequency $\omega_p\approx 1.37\times 10^{16}\,$rad/s
and the relaxation parameter at $T=300\,$K takes the value
$\gamma\approx 0.53\times 10^{14}\,$rad/s \cite{59}.

The dielectric permittivity of the plasma model, which disregards the dissipation
properties of conduction electrons, is obtained from (\ref{eq10}) by putting
$\gamma=0$

\begin{equation}
\ve_{\rm p}(\omega)=1-\frac{\omega_p^2}{\omega^2}, \qquad
\ve_{{\rm p},l}=1+\frac{\omega_p^2}{\xi_l^2}.
\label{eq11}
\end{equation}

\noindent
This model is physically applicable only at high frequencies in the region of
infrared optics. However, as mentioned in Section~1, the theoretical results
obtained using the plasma model at low frequencies, including the zero frequency,
agree with measurements of the Casimir force. As to the Drude model, which is
physically applicable at low frequencies, it leads to contradictions between
theoretical predictions of the Lifshitz theory and the measurement data.

As it was discussed many times in the literature starting from \cite{60,61},
the important formal difference between the dielectric permittivities (\ref{eq10})
and (\ref{eq11}) is that they lead to radically different values of the TE
reflection coefficient defined in (\ref{eq4}) at zero frequency

\begin{equation}
r_{\rm TE,D}(0,\kb)=0, \qquad
r_{\rm TE,p}(0,\kb)=
\frac{c\kb-\sqrt{c^2\skb+\omega_p^2}}{c\kb+\sqrt{c^2\skb+\omega_p^2}}.
\label{eq12}
\end{equation}

It immediately follows that at large separations, where the Casimir pressure
is determined by the terms of (\ref{eq7}) with $l=0$,

\begin{equation}
P_{{\rm TE,D}}^{\,0}(a,T)=0, \qquad
P_{{\rm TM,D}}^{\,0}(a,T)=\nrmP(a,T)=-\frac{k_BT}{8\pi a^3}\zeta(3),
\label{eq13}
\end{equation}

\noindent
where $\zeta(z)$ is the Riemann zeta function. This is one half of the result
obtained at large separations for the ideal metal planes.

For the plasma model, the case of ideal metal planes is obtained in the limit
$\omega_p\to\infty$ where

\begin{equation}
\lim_{\omega_p\to\infty} r_{\rm TE,p}(0,\kb)=-1
\label{eq14}
\end{equation}

\noindent
and the terms of (\ref{eq7}) with $l=0$ are

\begin{equation}
P_{{\rm TM},{\rm p}}^{\,0}(a,T)=
P_{{\rm TE},{\rm p}}^{\,0}(a,T)=-\frac{k_BT}{8\pi a^3}\zeta(3),
\qquad
P_{\!{\rm p}}^{\,0}(a,T)=-\frac{k_BT}{4\pi a^3}\zeta(3).
\label{eq15}
\end{equation}

\noindent
These are the same results as are obtained for the ideal metal planes. The quantities
(\ref{eq13}) and (\ref{eq15}) do not depend on $\hbar$. They represent the classical
limit of the Casimir pressure at large separations found using the Drude and plasma
models, respectively.

To determine the region of separations, where the dielectric permittivities of the
Drude and plasma models (\ref{eq10}) and (\ref{eq11}) determine nearly the total
Casimir pressure, we, first, compute the values of $P_{{\rm D}}$
and $P_{\!{\rm p}}$ and then compare the obtained results with the Casimir pressures
computed using the available optical data of Au extrapolated down to  zero
frequency by means of the Drude or plasma models.

\begin{figure}[H]
\vspace*{-7.2cm}
\centerline{\hspace*{-2.7cm}
\includegraphics[width=4.5in]{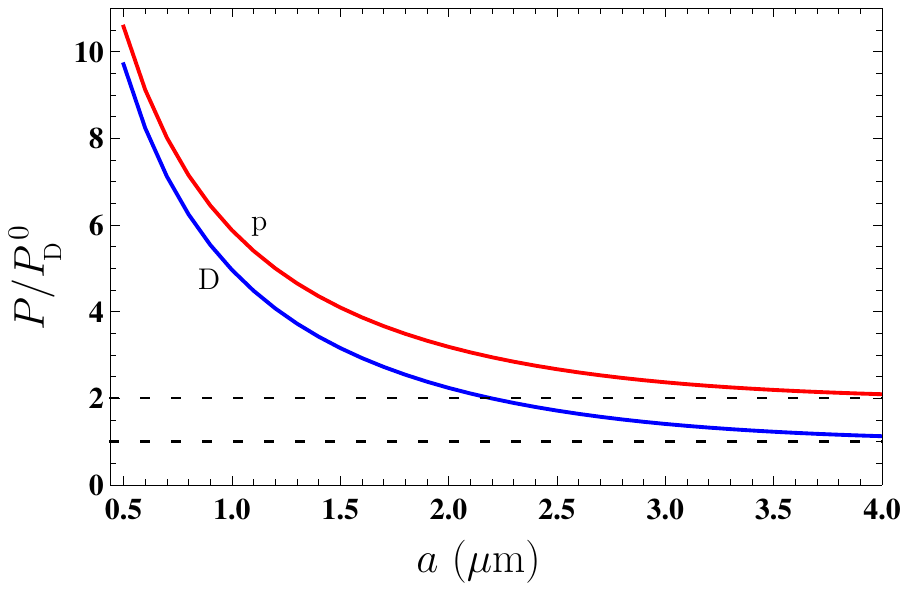}}
\vspace*{-3.7cm}
\caption{\label{fg1}
The ratio of the Casimir pressures for Au plates computed at $T=300\,$K
using the Drude or the plasma model to the classical limit of the Casimir
pressure found using the Drude model is shown as a function of separation
by the bottom and top lines, respectively.}
\end{figure}

Numerical computations of the Casimir pressure at $T=300\,$K were performed by
(\ref{eq6}) and (\ref{eq7}) with the reflection coefficients (\ref{eq4}) and the
dielectric permittivities (\ref{eq10}) and (\ref{eq11}). The computational results
for the ratios of obtained pressures to $\nrmP$ defined in (\ref{eq13}) are
presented in Figure~\ref{fg1} as a function of separation by the top and
bottom solid lines computed using the plasma and Drude models, respectively.
The two dashed lines indicate the corresponding limiting values of the pressure
ratios at large separations.

As is seen in Figure~\ref{fg1}, the theoretical predictions obtained using the plasma
and Drude models differ by the factors of 1.09 at $a=0.5\,\upmu$m, 1.2 at $a=1.1\,\upmu$m,
and 1.86 at $a=4\,\upmu$m. In the limit of large separations (classical limit) the
difference is by the factor of 2.

Now we determine the error in Casimir pressures made by omitting the contribution
of core electrons in the dielectric permittivity. For this purpose, we find the
dielectric permittivity of Au along the imaginary frequency axis by means of the
Kramers-Kronig relation where the imaginary part of this permittivity is given by
the tabulated optical data of Au \cite{59} extrapolated down to zero frequency by
means of the plasma or the Drude model (see, e.g., \cite{11,47} for details).
Then the Casimir pressure $P_{{{\rm D}},{\rm p}}^{\,\rm core}$ is again computed
by (\ref{eq4}), (\ref{eq6}) and (\ref{eq7}).

The relative deviation between the Casimir pressures obtained using the simple
Drude and plasma models and using the optical data taking into account the core
electrons can be characterized by the quantity

\begin{equation}
\delta P_{{\rm D,p}}(a,T)=
\frac{P_{{\rm D,p}}(a,T)-
P_{{\rm D,p}}^{\,\rm core}(a,T)}{P_{{\rm D,p}}^{\,\rm core}(a,T)}.
\label{eq16}
\end{equation}

\begin{figure}[H]
\vspace*{-7.2cm}
\centerline{\hspace*{-2.7cm}
\includegraphics[width=4.5in]{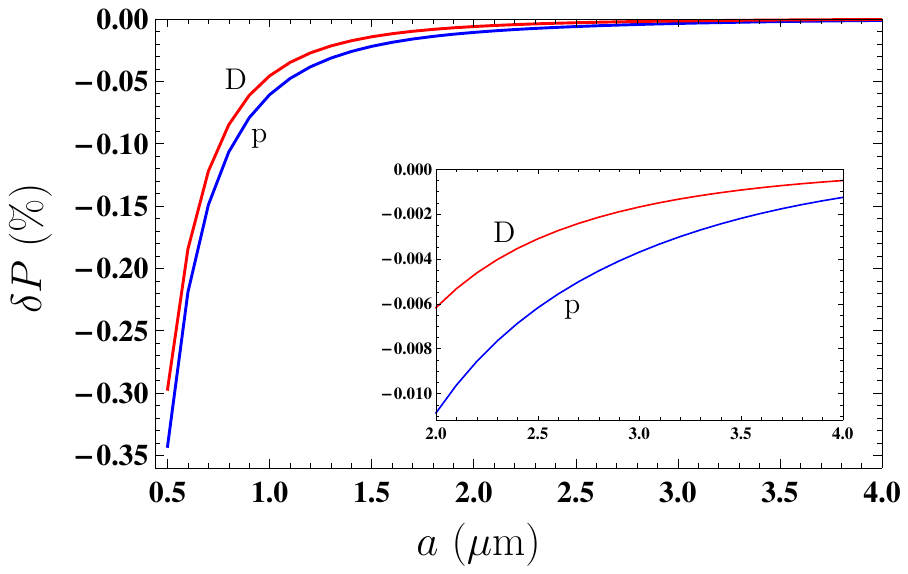}}
\vspace*{-3.7cm}
\caption{\label{fg2}
The relative deviation between the Casimir pressures for Au plates
computed at $T=300\,$K using the simple Drude or plasma model and
the optical data for Au extrapolated to zero frequency by the same
models is shown as a function of separation
by the top and bottom  lines, respectively. In the inset, the region of
large separations is shown on an enlarged scale.}
\end{figure}

In Figure~\ref{fg2}, the computational results for $\delta P_{{\rm D,p}}$
are shown as a function of separation by the top and bottom lines computed using
the Drude and plasma models and corresponding extrapolations of the optical data,
respectively.  In the inset, the region of separations from 2 to $4\,\upmu$m, where
the two lines are partially overlapping, is shown on an enlarged scale for better
visualization.

As is seen from Figure~\ref{fg2}, at $a=0.5\,\upmu$m the simple Drude and plasma
models reproduce the Casimir pressure computed with due regard for core electrons
with the relative errors less than 0.3\% and 0.35\%, respectively. These errors
quickly decrease with increasing separation. Thus, at $a=1\,\upmu$m they are below
0.05\% and 0.052\%, respectively.

Note that in the separation region above $0.5\,\upmu$m the already performed precision
determinations of the effective Casimir pressure between two parallel plates by
measuring the force gradient in the sphere-plate geometry
\cite{18,19,20,21,22,24,25,26,27,28,29,30} reliably distinguish between the top and
bottom lines in Figure~\ref{fg1} in favor of the former at $a<1.1\,\upmu$m.
However, at $a>0.5\,\upmu$m the same experiments cannot discriminate between the
theoretical predictions made by means of only the simple Drude or plasma model and
taking into account the core electrons.
As an example, the total experimental error in measuring the Casimir pressure
determined at the 67\% confidence level is $\delta P^{\rm expt}=1.5$\% at
$a=0.5\,\upmu$m \cite{20,21} and  $\delta P^{\rm expt}=27.5$\% at
$a=1.1\,\upmu$m \cite{29,30} (by measuring the Casimir force in the sphere-plate
geometry, the theoretical description using the Drude model was excluded at
all separations $a\leqslant 4.8\,\upmu$m \cite{23}).

\section{Comparison Studies of Contributions from the Propagating and
Evanescent Waves}

Now we are in a position to find all the four contributions to the Casimir
pressure (\ref{eq1}) when using the simple Drude and plasma models and determine
which of them is responsible for a disagreement between experiment and theory.
In accordance with the results of Section~3, this should be done at separations
between the plates exceeding $0.5\,\upmu$m where the dielectric permittivities
of the simple Drude and plasma models contribute nearly total value of the
pressure. There is no point also in considering too large separations because
the experimental situation there is uncertain. We begin with contribution
of the TM polarized waves to the Casimir pressure.

\subsection{Transverse Magnetic Polarization}

The contribution of the TM polarized waves, $P_{\rm TM}$, is calculated by
(\ref{eq7}) where the reflection coefficient $r_{\rm TM}$ is given by the
first equality in (\ref{eq4}) taken at $\omega=i\xi_l$. Depending on whether
we use the Drude (\ref{eq10}) or the plasma (\ref{eq11}) model of the
dielectric permittivity, we obtain either $P_{{\rm TM,D}}$ or
$P_{{\rm TM,p}}$.

\begin{figure}[H]
\vspace*{-7.6cm}
\centerline{\hspace*{-2.7cm}
\includegraphics[width=6.in]{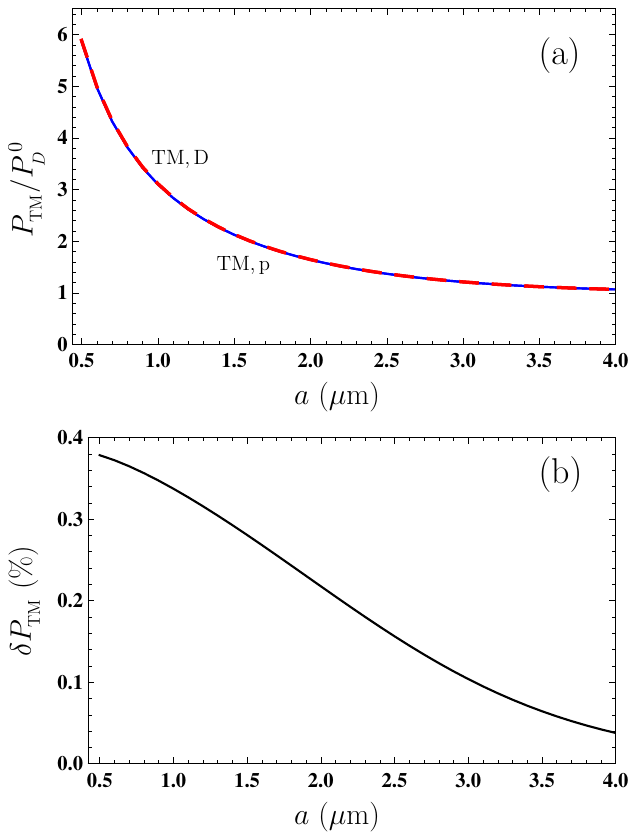}}
\vspace*{-3.8cm}
\caption{\label{fg3}
The transverse magnetic contribution to the Casimir pressure for Au plates
normalized to $\nrmP$ copmuted at $T=300\,$K using the simple Drude or
plasma model (a) and the relative deviation between these contributions (b)
are shown as a function of separation (a)
by the solid and dashed lines, respectively, and (b) by the solid line.}
\end{figure}

The computational results for $P_{\rm TM}$ normalized to $\nrmP$ at $T=300\,$K are
shown in Figure~\ref{fg3}(a) as a function of separation by the solid and dashed
lines computed using the Drude and plasma models, respectively. As can be seen in
this figure, the solid and dashed lines coincide very closely.

In order to understand the measure of agreement between the theoretical predictions of
the Lifshitz theory using the Drude and plasma  models we consider the relative
deviation

\begin{equation}
\delta P_{\rm TM}(a,T)=\frac{P_{{\rm TM,D}}(a,T)-
P_{{\rm TM, p}}(a,T)}{P_{{\rm TM,p}}(a,T)}.
\label{eq17}
\end{equation}

In Figure~\ref{fg3}(b), the computational results for $\delta P_{\rm TM}$ at
$T=300\,$K are shown by the solid line as a function of separation. As is seen
in this figure, the relative deviation between the predictions obtained using
these models decreases from approximately 0.38\% at $a=0.5\,\upmu$m to
 0.04\% at $a=4\,\upmu$m. Remembering that the Drude model takes into account
the dissipation processes, which are fully disregarded by the plasma model, one
can conclude that the transverse magnetic contribution to the Casimir pressure
between metallic plates is scarcely affected by the dissipation of conduction
electrons. It becomes clear also that the impact of dissipation in different
contributions to $P_{{\rm TM,D}}$ should be somehow compensated (see below).

Let us now determine the contributions of propagating and evanescent waves to
$P_{\rm TM}$ when using the Drude and plasma models in computations. The
contribution of evanescent waves is found by (\ref{eq3}) with the reflection
coefficient $r_{\rm TM}$ defined in (\ref{eq4}), whereas the contribution of
propagating waves can be obtained by (\ref{eq9}), where the total TM contribution
to the Casimir pressure is already computed [see Figure~\ref{fg3}(a)].

First of all, it is evident from (\ref{eq3}) that

\begin{equation}
P_{{\rm TM,p}}^{\,\rm evan}(a,T)=0.
\label{eq18}
\end{equation}

\noindent
This is because the dielectric permittivity of the plasma model (\ref{eq11})
and, thus, the reflection coefficient $r_{\rm TM,p}$ in (\ref{eq4}) are the real
functions for evanescent waves.

Then, from (\ref{eq9}) one concludes that

\begin{equation}
P_{{\rm TM,p}}^{\,\rm prop}(a,T)=
P_{\,{\rm TM,p}}(a,T),
\label{eq19}
\end{equation}

\noindent
where $P_{{\rm TM,p}}$ is already shown by the red dashed line
in Figure~\ref{fg3}(a).

For the Drude model, the computations of $P_{{\rm TM,D}}^{\,\rm evan}$
are again performed   by (\ref{eq3}) with the reflection
coefficient $r_{\rm TM,D}$ defined in (\ref{eq4}) and the dielectric permittivity
(\ref{eq10}). The quantity $P_{{\rm TM,D}}^{\,\rm prop}$  is obtained
from (\ref{eq9}) where the already computed $P_{{\rm TM,D}}$ is shown by
the solid line in  Figure~\ref{fg3}(a).

\begin{figure}[H]
\vspace*{-7.2cm}
\centerline{\hspace*{-2.7cm}
\includegraphics[width=4.5in]{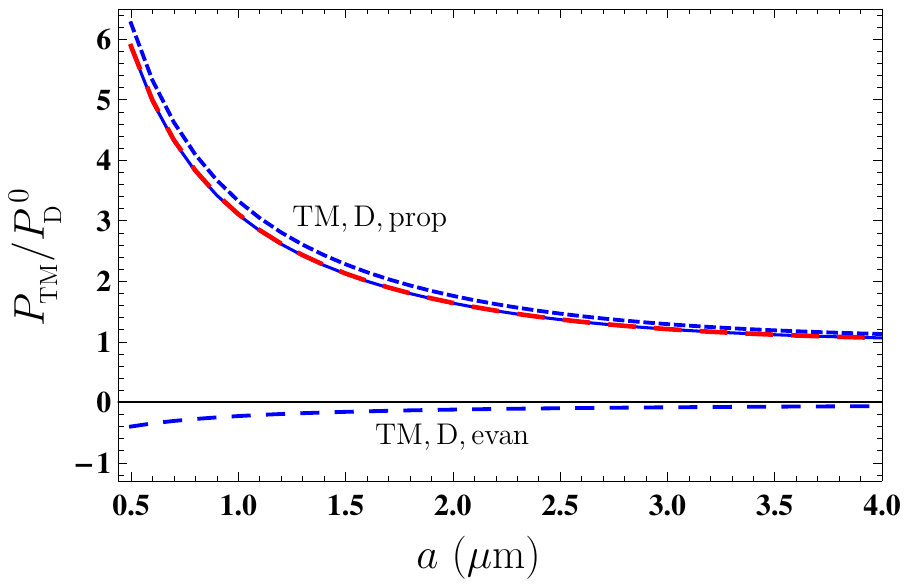}}
\vspace*{-3.7cm}
\caption{\label{fg4}
The transverse magnetic contributions to the Casimir pressure for Au plates
due to propagating and evanescent waves
normalized to $\nrmP$ computed at $T=300\,$K using the simple Drude
model are shown as a function of separation
by the top short-dashed and bottom long-dashed blue lines, respectively.
The solid blue and long-dashed red lines for the normalized total transverse
magnetic contributions to the Casimir pressure computed using the Drude and
plasma models are reproduced from Figure~\ref{fg3}(a).}
\end{figure}

Figure~\ref{fg4} shows the computational results for $P_{{\rm TM,D}}^{\,\rm prop}$
and $P_{{\rm TM,D}}^{\,\rm evan}$ at $T=300\,$K by the top short-dashed and
bottom long-dashed lines as a function of separation. Both these lines are blue.
For comparison purposes, in Figure~\ref{fg4} we also reproduce from Figure~\ref{fg3}(a)
the blue solid line and the overlapping with it red dashed line demonstrating the
separation dependence of $P_{{\rm TM,D}}$ and $P_{{\rm TM,p}}$,
respectively (the latter also depicts the behavior of
$P_{{\rm TM,p}}^{\,\rm prop}$).

{}From Figure~\ref{fg4} it is seen that, although the quantities $P_{\rm TM,D}$
and $P_{\rm TM,p}$ are almost equal, their constituent parts due to the
propagating and evanescent waves are different. For the plasma model,
$P_{\rm TM,p}$ is determined entirely by the propagating waves, whereas for the
Drude model the contribution of
$P_{{\rm TM,D}}^{\,\rm prop}$ to $P_{\rm TM,D}$  is partially compensated
by $P_{{\rm TM,D}}^{\,\rm evan}$ which is of the opposite sign, i.e.,
corresponds to the Casimir repulsion. This explains why there is no eventual
impact of dissipation on $P_{\rm TM,D}$, even though the Drude model is dissipative.

\subsection{Transverse Electric Polarization}

\begin{figure}[H]
\vspace*{-7.2cm}
\centerline{\hspace*{-2.7cm}
\includegraphics[width=4.5in]{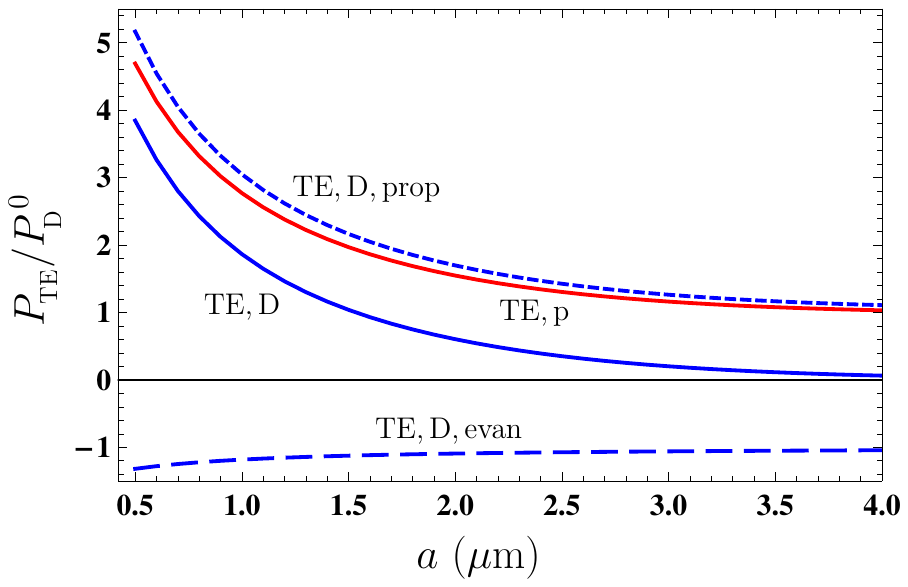}}
\vspace*{-3.7cm}
\caption{\label{fg5}
The transverse electric contributions to the Casimir pressure for Au plates
due to propagating and evanescent waves
normalized to $\nrmP$ computed at $T=300\,$K using the simple Drude
model and the total transverse electric contribution
are shown as a function of separation
by the top and bottom short-dashed,  long-dashed lines, and the
lower solid line, respectively.
The upper solid line shows similar results for the transverse electric
contribution computed using the simple plasma model.}
\end{figure}

We calculate the contribution of the transverse electric polarization, $P_{\rm TE}$,
to the Casimir pressure by (\ref{eq7}) with the reflection coefficient $r_{\rm TE}$
from (\ref{eq4}) using the dielectric permittivities of the Drude model (\ref{eq10})
and the plasma model (\ref{eq11}).
In Figure~\ref{fg5}, the computational results for $P_{\rm TE,D}$ and $P_{\rm TE,p}$
normalized to $\nrmP$
at $T=300\,$K are shown
as a function of separation by the lower (blue) and upper (red) solid lines for
the Drude and plasma models, respectively.

{}From Figure~\ref{fg5} it is seen that the lower and upper solid lines differ
considerably. Keeping in mind that, according to the results of Section~4.1,
$P_{{\rm TM,D}}$ and $P_{{\rm TM},{\rm p}}$ are equal with a high
degree of accuracy, it becomes clear that this difference completely determines
the discrepancy between the total Casimir pressures computed using the Drude
and plasma models, $P_{\rm D}$ and $P_{\rm p}$. The question arises what is
the physical origin of this discrepancy.

To answer this question, we compute the quantities $P_{{\rm TE,D}}^{\,\rm evan}$
and $P_{{\rm TE},{\rm p}}^{\,\rm evan}$ by (\ref{eq3}). As to the latter, it is
evident that

\begin{equation}
P_{{\rm TE},{\rm p}}^{\,\rm evan}(a,T)=0,
\label{eq20}
\end{equation}

\noindent
because the dielectric permittivity of the plasma model (\ref{eq11})
and the reflection coefficient $r_{\rm TE,p}$ from (\ref{eq4}) are the real
functions in the region of evanescent waves.

Taking into account (\ref{eq9}), we also find that

\begin{equation}
P_{{\rm TE},{\rm p}}^{\,\rm prop}(a,T)=
P_{{\rm TE},{\rm p}}(a,T),
\label{eq21}
\end{equation}

\noindent
i.e., that for the plasma model the total Casimir pressure determined by the
transverse electric polarization is equal to the contribution of TE-polarized
propagating waves. This is the same as was proven above for the TM polarization.
Thus, $P_{{\rm TE},{\rm p}}^{\,\rm prop}$ is given by the upper solid line
in Figure~\ref{fg5}(a) already drawn for $P_{{\rm TE},{\rm p}}$.

The computational results for $P_{{\rm TE,D}}^{\,\rm evan}$ obtained by
(\ref{eq3}), (\ref{eq4}) and (\ref{eq10}) at $T=300\,$K are shown as a function
of separation in Figure~\ref{fg5} by the bottom long-dashed line.
As to the computational results for $P_{{\rm TE,D}}^{\,\rm prop}$, they are
found from (\ref{eq9}) and shown by the top short-dashed line in Figure~\ref{fg5}
as a function of separation.

All contributions to $P_{\rm TE}$ are now computed using both models of the
dielectric response of Au and it is possible to analyze the role of each of them.
First of all, from Figure~\ref{fg5} it is seen that the deviation between
$P_{{\rm TE,D}}^{\,\rm prop}$ and $P_{{\rm TE},{\rm p}}^{\,\rm prop}$
shown by the top short-dashed line and the upper solid line, respectively
(we recall that the latter line also shows $P_{{\rm TE},{\rm p}}$), is
reasonably small and cannot be responsible for a much larger discrepancy
between $P_{\rm D}$ and $P_{\rm p}$. The latter is equal to the discrepancy
between $P_{{\rm TE,D}}$ and $P_{{\rm TE},{\rm p}}$ shown by the two
solid lines. In fact, the deviation between
$P_{{\rm TE,D}}^{\,\rm prop}$ and $P_{{\rm TE},{\rm p}}^{\,\rm prop}$
demonstrates the impact of dissipation of conduction electrons on the TE
contribution to the Casimir pressure, which is taken into account by the Drude
model and disregarded by the plasma one. It is significant that this impact carried
out through the TE propagating waves is not in contradiction with the experimental
data on measuring the Casimir force.

A completely different type of situation occurs for $P_{{\rm TE,D}}^{\,\rm evan}$
shown by the bottom long-dashed line in Figure~\ref{fg5}. The magnitude of
$P_{{\rm TE,D}}^{\,\rm evan}$  is much larger than
$P_{{\rm TM,D}}^{\,\rm evan}$,  and this leads to a significant deviation
between  $P_{{\rm TE,D}}$ and $P_{{\rm TE},{\rm p}}$ resulting ultimately
in a contradiction between the measurement data and theoretical predictions of the
Lifshitz theory obtained using the Drude model.

For better understanding of this situation, we should take into account that the
Drude model has a wealth of alternative experimental confirmations in the area of
propagating waves with any polarization, as well as for the transverse magnetic
evanescent waves, but lacks of confirmation for the transverse electric evanescent
waves (see a discussion of experimental situation in the next section). On this basis
one can conclude that experiments on measuring the Casimir force between metallic
test bodies invalidate the dielectric permittivity of the Drude model in the area
of transverse electric evanescent waves.  It is apparent that the alternative experimental
confirmations of such a conclusion are highly desirable (see the next section).

\section{Discussion: Failure of the Drude Model for Transverse Electric
Evanescent Waves, the Role of Dissipation, and Possibilities of Alternative
Tests}

As discussed in Section 1, the theoretical predictions of the fundamental
Lifshitz theory are in conflict with the measurement data of many precision
experiments of Casimir physics if the dielectric response of conduction
electrons is described by the dissipative Drude model. However, by
disregarding the dissipation properties of conduction electrons, i.e., by
using the plasma model, one can bring the measurement data in agreement
with the theoretical predictions. This situation is unacceptable because
the dissipation of conduction electrons at low frequencies is the much
studied and confirmed by many experiments physical effect.

According to the results presented above, an account of dissipation by
means of the Drude model in the transverse magnetic contribution to the
Casimir pressure leads to the same results as are obtained using the
dissipationless plasma model. This is because the dissipation-induced
terms in the Casimir pressure arising from the evanescent and propagating
waves cancel each other. The dissipation-induced term in the contribution
to the Casimir pressure from the transverse electric propagating waves is
found to be reasonably small and does not bring the theoretical predictions
found using the Drude model in contradiction with the measurement data.

The performed computations show that the roots of contradiction are not
in the account of dissipation in itself, but in how the Drude model
describes the response of metals to the low-frequency transverse electric
evanescent waves. These computations compared with the measurement data
lead us to conclude that the theoretical description of the electromagnetic
response of metals to the transverse electric evanescent waves given by the
Drude model is in error. In this context, it is necessary to discuss what
are the alternative experimental evidences about the validity of the Drude
model other than the Casimir effect.

In the area of both the transverse magnetic and transverse electric propagating
waves there is an abundance of experimental confirmations of the Drude
model in physics, electrotechnics, and even in the day-to-day life, so that
it makes no sense to discuss them. However, direct measurement of the
reflection coefficients of a metal in the case of evanescent waves presents
difficulties because all commonly used methods (ellipsometry, for instance)
are adapted for the propagating waves.

Great interest to the evanescent waves during the last decades is connected
with the fact that they made it possible to surmount the optical
diffraction limit \cite{62}. Thus, the physics of plasmons polaritons gives
the possibility to obtain the great deal of evidence about the reflection
of evanescent waves on metallic surfaces, but only for the transverse
magnetic polarization \cite{63}. The reflectivity properties of weakly
evanescent waves (for which $\kb$ is only just above $\omega/c$) can
be examined by means of the total internal reflection and frustrated total
internal reflection \cite{64,65,66}. The widely used in various
technological applications near-field optical microscopy is reasonably
sensitive to only the transverse magnetic evanescent waves \cite{67,67a}
(see also the discussion in \cite{58} for more details).

The information provided above allows to conclude that the failure of the
Drude model demonstrated by experiments on measuring the Casimir force
does not contradict to all the available experimental evidences in favor
of this model which are irrelevant to the area of transverse electric
evanescent waves.

Despite the fact that there are many experiments mentioned above, which
demonstrate the failure of the Drude model resulting from the region of
transverse electric evanescent waves, it would be highly desirable to
perform one more independent test disconnected with the Casimir effect.
Recently such an alternative test in the field of classical
electrodynamics was proposed in \cite{58,68}. It was shown that the
lateral components of magnetic field of an oscillating magnetic dipole
spaced in the proximity of metallic plate are determined by solely the
transverse electric evanescent waves. According to the results of
\cite{58,68}, by choosing the suitable dipole frequency and using either
the Drude or the plasma model for the dielectric permittivity of metallic
plate, the lateral components of the dipole field are varied by up to
several orders of magnitide depending on the model used. Thus, by
measuring these components for some fixed dipole parameters, it is
possible to reliably conclude whether the Drude model describes correctly
the response of plate metal to the transverse electric evanescent waves.

As an example, in \cite{58,68} the magnetic dipole of 1 mm size with the
dipole moment of $3.14\times 10^{-5}\,\mbox{Am}^2$ oscillating with the frequency of
100~rad/s at 1~cm height above the Cu plate was considered. Small dipoles of
such kind are manufactured in the form of coils containing of about 10
turns \cite{69,70,71}. In this case, the lateral component of the dipole
magnetic field at the same height of 1~cm above the plate computed using
the Drude model was found to be $0.027\,\mbox{A/m}=3.37\times 10^{-8}\,$T \cite{58,68}.
If the plasma model is used in computations, the larger by a factor of
10 magnetic field is obtained \cite{58,68}. Keeping in mind that the
current resolution limit in measurements of weak magnetic fields is of
about $10^{-13}$ T \cite{72,73,74}, the proposed alternative test of
the Drude model in the region of transverse electric evanescent waves
seems quite feasible.

Finally, if it is confirmed that the Drude model is really invalid in
the region of low-frequency transverse electric evanescent waves, the
question arises on how it could be corrected. Recently the modifications
of the Drude model at low frequencies caused by the spatial dispersion
were again considered \cite{75,76} in connection with the problems of
Casimir physics. The suggested modifications, however, are incapable
to bring the theoretical predictions in agreement with the measurement
data for the Casimir force. The phenomenological spatially nonlocal
alterations in the Drude model, which bring the theoretical predictions
in agreement with all performed experiments on measuring the Casimir
force, were suggested in \cite{77,78,79}, but they are still lacking of
fundamental theoretical justification. Thus, the proper form of
the response function of metals to the transverse electric evanescent
waves remains to be found.

\section{Conclusions}

To conclude, in this paper we have performed the comparison studies of
four contributions to the Casimir pressure between metallic plates
caused by the transverse magnetic and transverse electric polarizations
of the electromagnetic field and by the propagating and evanescent
waves. The region of separations was determined where the major
contribution to the pressure is given by the electromagnetic response
of free electrons described by the dissipative Drude model or the
experimentally consistent but dissipationless plasma model used in
comparisons between experiment and theory.

According to our results, the transverse magnetic contributions to the
Casimir pressure computed by using the Drude or plasma models are equal
to a high degree of accuracy. In so doing, if the Drude model is used,
the relatively small contribution from the evanescent waves (which is
equal to zero when using the plasma model) is canceled by an excessive
contribution from the propagating waves. Thus, the use of the Drude model
for computation of the Casimir pressure determined by the transverse
electric polarization does not lead to contradictions between experiment
and theory.

It was shown also that the transverse electric contribution to the
Casimir pressure caused by the propagating waves, which is computed by
using the Drude model, deviates reasonably small from the transverse
electric contribution computed using the plasma model (the latter is
again determined by the propagating waves alone). This deviation is due
to the dissipation processes of propagating waves taken into account
by the Drude model. It cannot explain a discrepancy between the
theoretical predictions obtained using the Drude model and the
measurement data because of its smallness.

Next, it was found that the experimental inconsistency of the Drude
model is determined by the relatively large contribution of the transverse
electric evanescent waves. This leads to a conclusion that the response
of metals to the transverse electric evanescent waves is described by the
Drude model incorrectly. In such a manner, the reason why the Lifshitz
theory using the Drude model is experimentally inconsistent is not that
it takes into account  dissipation of free electrons, as opposed to
the plasma model, but that it takes it into account incorrectly in the
region of the transverse electric evanescent waves.

The presented analysis of experimental tests of the Drude model
demonstrates that it is lacking experimental confirmation in this
important region of the wave vectors and frequencies. Therefore, the
recently proposed alternative test of the Drude model as a response
function to the transverse electric evanescent waves should shed new
light on the problem of disagreement between theoretical predictions
of the Lifshitz theory and the measurement data.

\vspace{6pt}

\funding{G.L.K. was partially funded by the
Ministry of Science and Higher Education of Russian Federation
("The World-Class Research Center: Advanced Digital Technologies",
contract No. 075-15-2022-311 dated April 20, 2022). The research
of V.M.M. was partially carried out in accordance with the Strategic
Academic Leadership Program "Priority 2030" of the Kazan Federal
University. }

\begin{adjustwidth}{-\extralength}{0cm}

\reftitle{References}

\end{adjustwidth}
\end{document}